\documentclass[a4paper, 11pt]{article} % for Physics and Applied Mathematics and Statistics articles
\pdfoutput=1

\usepackage{jheppub}
\usepackage{url,hyperref,lineno,microtype,subcaption}
\usepackage[onehalfspacing]{setspace}
\usepackage[ruled,vlined]{algorithm2e}

\usepackage{enumerate}
\usepackage{tabularx}
\usepackage{lscape}

\author[a]{Marco Rovere}
\author[b]{Ziheng Chen}
\author[c,d]{Antonio Di Pilato}
\author[a]{Felice Pantaleo}
\author[e]{Chris Seez}
\affiliation[a]{European Organization for Nuclear Research (CERN), Esplanade des
Particules 1, 1211 Meyrin, Switzerland}
\affiliation[b]{Northwestern University, 633 Clark Street, Evanston, IL 60208,
US}
\affiliation[c]{University of Bari, Piazza Umberto I, 1, 70121 Bari, Italy}
\affiliation[d]{National Institute for Nuclear Physics (INFN) - Sezione di Bari,
Via Amendola 173, 70126 Bari, Italy}
\affiliation[e]{Imperial College London, South Kensington Campus, London SW7 2AZ, UK}

\emailAdd{marco.rovere@cern.ch}
\emailAdd{zihengchen2015@u.northwestern.edu}
\emailAdd{antonio.dipilato@ba.infn.it}
\emailAdd{felice.pantaleo@cern.ch}
\emailAdd{chris.seez@cern.ch}

\title{CLUE: A Fast Parallel Clustering Algorithm for High Granularity Calorimeters in High Energy Physics}
\abstract{
One of the challenges of high granularity calorimeters, such as that to be built
to cover the endcap region in the CMS Phase-2 Upgrade for HL-LHC, is that the
large number of channels causes a surge in the computing load when clustering
numerous digitised energy deposits (hits) in the reconstruction stage. In this
article, we propose a fast and fully-parallelizable density-based clustering
algorithm, optimized for high occupancy scenarios, where the number of clusters
is much larger than the average number of hits in a cluster. The algorithm uses
a grid spatial index for fast querying of neighbours and its timing scales
linearly with the number of hits within the range considered. We also show a
comparison of the performance on CPU and GPU implementations, demonstrating the
power of algorithmic parallelization in the coming era of heterogeneous
computing in high energy physics.

}

\begin{document}
\maketitle

%---------------------
% sections
%---------------------
\section{Introduction}
\label{sec:introduction}

Calorimeters with high lateral and longitudinal readout granularity, capable of
providing a fine grained image of electromagnetic and hadronic showers, have
been suggested for future high energy physics experiments
\cite{calice2012calorimetry}. The  silicon sensor readout cells of the CMS
endcap calorimeter (HGCAL) \cite{Collaboration:2293646} for HL-LHC
\cite{Apollinari:2284929} have an area of about $1~\mathrm{cm}^2$.  When a
particle showers, the deposited energy is collected by the sensors on the layers
which the shower traverses.  The purpose of the clustering algorithm when
applied to shower reconstruction is to group together individual energy deposits
(hits) originating from a particle shower. Due to the high lateral granularity,
the number of hits per layer is large, and it is computationally advantageous to
collect together hits in 2D clusters layer-by-layer \cite{Chen:2017btc} and then
associate these 2D clusters, representing energy blobs, in different layers
\cite{Collaboration:2293646}.

However, a computational challenge emerges as a consequence of the large data
scale and limited time budget. %For example, clustering millions of hits in each
event is tightly constrained by a millisecond-level execution time.  This
constraint requires the clustering algorithm to be highly efficient while
maintaining a low computational complexity. Furthermore, a linear scalability is
strongly desired in order to avoid bottlenecking the performance of the entire
event reconstruction.  Finally, it is highly preferable to have a
fully-parallelizable clustering algorithm to take advantage of the trend of
heterogeneous computing with hardware accelerators, such as graphics processing
units (GPUs), achieving a higher event throughput and a better energy
efficiency.

% input/output, characteristics
The input to the clustering algorithm is a set of $n$ hits, whose number varies
from a few thousands to a few millions, depending on the longitudinal and
transverse granularity of the calorimeter as well as on the number of particles
entering the detector. The output is a set of $k$ clusters whose number is
usually one or two order of magnitudes smaller than $n$ and in principle depends
on both the number of incoming particles and the number of layers. Assuming that
the lateral granularity of sensors is constant and finite, the average number of
hits in clusters ($m=n/k$) is also constant and finite. For example, in the CMS
HGCAL, $m$ is in the order of 10. This leads to the relation among the number of
hits $n$, the number of clusters $k$, and the average number of hits in clusters
$m$ as $n > k \gg m$.

Most well-known algorithms do not simultaneously satisfy the requirements on
linear scalability and easy parallelization for applications like clustering
hits in high granularity calorimeters, which is characterized by low dimension
and $n > k \gg m$. It is therefore important to investigate new, fast and
parallelizable clustering algorithms, as well as their optimized accompanying
spatial index that can be conveniently constructed and queried in parallel.

In this paper, we describe CLUE (CLUstering of Energy), a novel and parallel
density-based clustering. Its development was inspired by the work described in
ref.~\cite{rodriguez2014clustering}. In Section~\ref{sec:algorithm}, we describe
the CLUE algorithm and its accompanying spatial index. Then in
Section~\ref{sec:implementation}, some details of GPU implementations are
discussed. Finally, in Section~\ref{sec:performance} we present CLUE's ability
on non-spherical cluster shapes and noise rejection, followed by its
computational performance when executed on CPU and GPU with synthetic data,
mimicking hits in high granularity calorimeters.

\section{Clustering Algorithm}
\label{sec:algorithm}

% review of current popular method
Clustering data is one of the most challenging tasks in several scientific domains. The definition of cluster is itself not trivial, as it strongly depends on the context. Many clustering methods have been developed based on a variety of induction principles \cite{maimon2005data}. Currently popular clustering algorithms include (but are not limited to) partitioning, hierarchical and density-based approaches \cite{maimon2005data,han2011data}. Partitioning approaches, such as k-mean \cite{lloyd1982least}, compose clusters by optimizing a dissimilarity function based on distance. However, in the application to high granularity calorimeters, partitioning approaches are prohibitive because the number of clusters $k$ is not known a priori. Hierarchical methods make clusters by constructing a dendrogram with a recursion of splitting or merging. However, hierarchical methods do not scale well because each decision to merge or split needs to scan over many objects or clusters \cite{han2011data}. Therefore, they are not suitable for our application. Density-based methods, such as DBSCAN \cite{Ester:1996:DAD:3001460.3001507}, OPTICS \cite{Ankerst:1999:OOP:304182.304187} and Clustering by Fast Search and Find Density Peak (CFSFDP) \cite{rodriguez2014clustering}, group points by detecting continuous high-density regions. They are capable of discovering clusters of arbitrary shapes and are efficient for large spatial database. If a spatial index is used, their computational complexity is $O(n\log n)$ \cite{han2011data}. However, one of the potential weaknesses of the currently well-known density-based algorithms is that they intrinsically include serial processes which are hard to parallelize: DBSCAN has to iteratively visit all points within an enclosure of density-connectedness before working on the next cluster \cite{Ester:1996:DAD:3001460.3001507}; OPTICS needs to sequentially add points in an ordered list to obtain a dendrogram of reachability distance \cite{Ankerst:1999:OOP:304182.304187}; CFSFDP needs to sequentially assign points to clusters in order of decreasing density \cite{rodriguez2014clustering}. In the application to high granularity calorimeters, as discussed in Section~\ref{sec:introduction}, linear scalability and fully parallelization are essential to handle a huge dataset efficiently by means of heterogeneous computing.

% our method
In order to satisfy these requirements, we propose a fast and fully-parallelizable density-based algorithm (CLUE) inspired by CFSFDP. For the purpose of the algorithm, each sensor cell on a layer with its energy deposit is taken as a 2D point with an associated weight equalling to its energy value. As in CFSFDP, two key variables are calculated for each point: the local density $\rho$ and the separation $\delta$ defined in Equation~\ref{eqn:algorithm:defineRho} and \ref{eqn:algorithm:defineDelta}, where $\delta$ is the distance to the nearest point with higher density (``nearest-higher'') which is slightly adapted from that in CFSFDP in order to take advantage of the spatial index. Then cluster seeds and outliers are identified based on thresholds on $\rho$ and $\delta$. Differing from cluster assignment in CFSFDP, which sorts density and adds points to clusters in order of decreasing density, CLUE first builds a list of followers for each point by registering each point as a follower to its nearest-higher. Then it expands clusters by passing cluster indices from the seeds to their followers iteratively. Since such expansion of clusters is fully independent from each others, it not only avoids the costly density sorting in CFSFDP, but also enables a $k$-way parallelization. Unlike the noise identification in CFSFDP, CLUE rejects noise by identifying outliers and their iteratively descendant followers, as discussed in Section~\ref{sec:performance:clusteringResults}.

%%%%%%%%%%%%%%%%%%%%%
% Spatial Index with Grid
%%%%%%%%%%%%%%%%%%%%%

\subsection{Spatial index with fixed-grid}
Query of neighbourhood, which retrieves nearby points within a distance, is one of the most frequent operations in density-based clustering algorithms. CLUE uses a spatial index to access and query spatial data points efficiently. Given that the physical layout of sensor cells is a multi-layer tessellation, it is intuitive to index its data with a fixed-grid, which divides the space into fixed rectangular bins \cite{bentley1979data,levinthal1966molecular}. Comparing with the data-driven structures such as KD-Tree \cite{Bentley:1975:MBS:361002.361007} and R-Tree \cite{Guttman:1984:RDI:971697.602266}, space partition in fixed-grid is independent of any particular distribution of data points \cite{rigaux2001spatial}, thus can be explicitly predefined before loading data points. In addition, both construction and query with a fixed-grid are computationally simple and can be easily parallelized. Therefore, CLUE uses a fixed-grid as spatial index for efficient neighbourhood queries.

% search box
\begin{figure}[ht]
    \centering
    \includegraphics[width=0.4\textwidth]{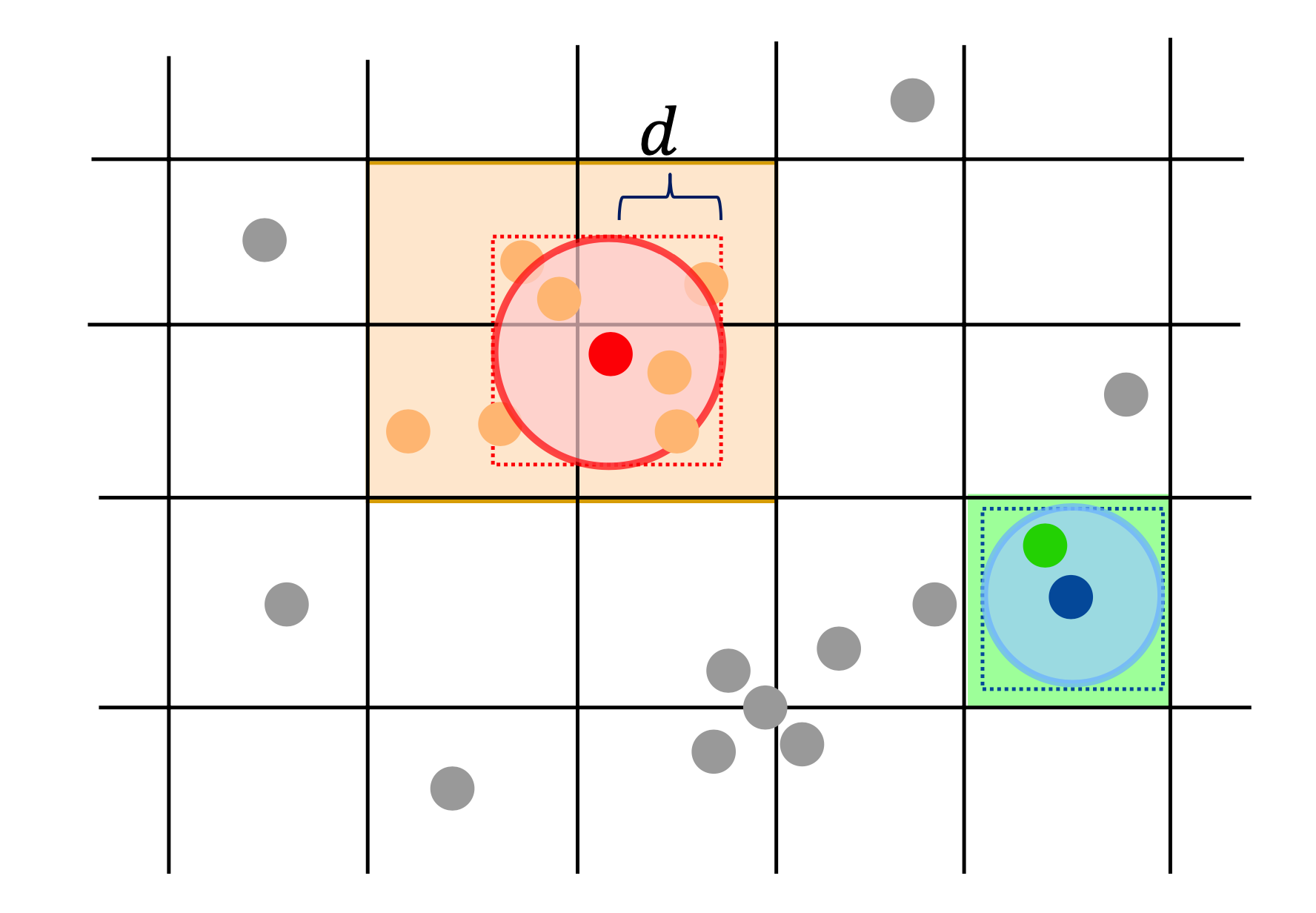}
    \caption{2D points are indexed with a grid for fast neighbourhood query in CLUE. Construction of this spatial index only involves registering the indices of points into the bins of the grid according to points' 2D spatial positions. To query d-neighbourhood $N_d(i)$ defined in Equation~\ref{eqn:algorithm:defineNeighborhood}, taking the red (blue) point for example, we first locate its $\Omega_d(i)$ defined in Equation~\ref{eqn:algorithm:defineSearchBox}, a set of all points in the bins touched by a square window $[x_i\pm d,y_i\pm d]$. The $[x_i\pm d,y_i\pm d]$ window is shown as the orange (green) square while $\Omega_d(i)$ is shown as orange (green) points. Then we examine points in $\Omega_d(i)$ to identify those within a distance $d$ from point $i$, shown as the ones contained in the red (blue) circle.}
    \label{fig:algorithm:searchBox}
\end{figure}

For each layer of the calorimeter, a fixed-grid spatial index is constructed by registering the indices of 2D points into the square bins in the grid according to the 2D coordinates of the points. When querying $N_d(i)$, the d-neighbourhood of point $i$, CLUE only needs to loop over points in the bins touched by the square window $(x_i\pm d,y_i\pm d)$ as shown in Fig.~\ref{fig:algorithm:searchBox}. We denote those points as $\Omega_d(i)$, defined as:
\begin{equation} \label{eqn:algorithm:defineSearchBox}
    \Omega_d(i) = \{j : j \in \text{tiles touched by the square window } [x_i\pm d,y_i\pm d] \}.
\end{equation}
\noindent where $\Omega_d(i)$ is guaranteed to include all neighbours within a distance $d$ from the point $i$. Namely,
\begin{equation} \label{eqn:algorithm:defineNeighborhood}
    N_d(i)=\{j: d_{ij}<d, j \in \Omega_d(i) \} \subseteq \Omega_d(i).
\end{equation}
\noindent Without any spatial index, the query of $N_d(i)$ requires a sequential scan over all points. In contrast, with the grid spatial index, CLUE only needs to loop over the points in $\Omega_d(i)$ to acquire $N_d(i)$. Given that $d$ is small and the maximum granularity of points is constant, the complexity of querying $N_d(i)$ with a fixed-grid is $O(1)$.

%%%%%%%%%%%%%%%%%%%%%
% Parameters and Procedure
%%%%%%%%%%%%%%%%%%%%%

\subsection{Clustering procedure of CLUE}

CLUE requires the following four parameters: $d_c$ is the cut-off distance in the calculation of local density; $\rho_c$ is the minimum density to promote a point as a seed or the maximum density to demote a point as an outlier; $\delta_c$ and $\delta_o$ are the minimum separation requirements for seeds and outliers, respectively. The choice of these four parameters can be based on physics: for example, $d_c$ can be chosen based on the shower size and the lateral granularity of detectors; $\rho_c$ can be chosen to exclude noise; $\delta_c$ and $\delta_o$ can be chosen based on the shower sizes and separations. These four parameters allow more degrees of freedom to tune CLUE for the desired goals of physics. %, such as better energy resolution and better pile-up rejection.

\begin{figure}[ht]
    \centering
    \includegraphics[trim=5cm 0cm 4cm 0cm, clip,width=0.99\textwidth]{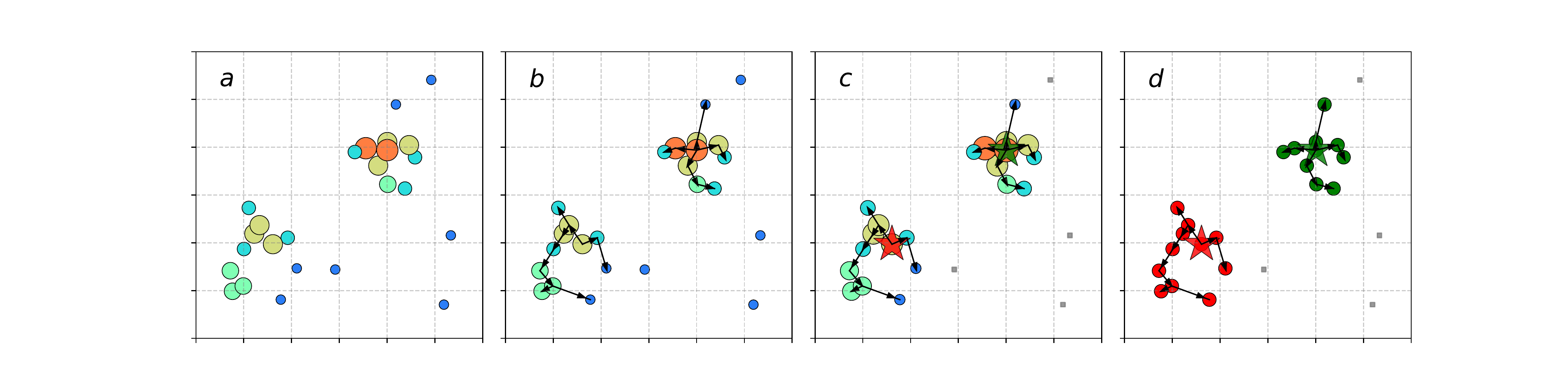}
    \caption{Demonstration of CLUE algorithm. Points are distributed inside a $6\times6$ 2D area and CLUE parameters are set to $d_c=0.5,\rho_c=3.9,\delta_c=\delta_o=1$. Before the clustering procedure starts, a fixed-grid spatial index is constructed. In the first step, shown as Fig.~\ref{fig:algorithm:procedure} (a), CLUE calculates the local density $\rho$ for each point, which is defined in Equation~\ref{eqn:algorithm:defineRho}. The color and size of points represent their local densities. In the second step, shown as Fig.~\ref{fig:algorithm:procedure} (b), CLUE calculates the nearest-higher $nh$ and the separation $\delta$ for each point, which are defined in Equation~\ref{eqn:algorithm:defineDelta}. The black arrows represent the relation from the nearest-higher of a point to the point itself. If the nearest-higher of a point is -1, there is no arrow pointing to it. In the third step, shown as Fig.~\ref{fig:algorithm:procedure} (c), CLUE promotes a point as a seed if $\rho,\delta$ are both large, or demote it to an outlier if $\rho$ is small and $\delta$ is large. Promoted seeds and demoted outliers are shown as stars and grey squares, respectively. In the fourth step, shown as Fig.~\ref{fig:algorithm:procedure} (d), CLUE propagates the cluster indices from seeds through their chains of followers defined in Equation~\ref{eqn:algorithm:defineFollowers}. Noise points, which are outliers and their descendant followers, are guaranteed not to receive any cluster ids from any seeds. The color of points represents the cluster ids. A grey square means its cluster id is undefined and the point should be considered as noise.
    }
    \label{fig:algorithm:procedure}
\end{figure}

% rho
Figure~\ref{fig:algorithm:procedure} illustrates the main steps of CLUE algorithm. The local density $\rho$ in CLUE is defined as:
\begin{equation} \label{eqn:algorithm:defineRho}
    \rho_i = \sum_{j: j \in N_{d_c}(i)} \chi(d_{ij}) w_j,
\end{equation}
\noindent where $w_j$ is the weight of point $j$, $\chi(d_{ij})$ is a convolution kernel, which can be optimized according to specific applications. Obvious possible kernel options include flat, Gaussian and exponential functions.

% delta
The nearest-higher and the distance to it $\delta$ (separation) in CLUE are defined as:
\begin{equation} \label{eqn:algorithm:defineDelta}
    nh_i =
    \begin{cases}
        \arg\min_{j \in N'_{d_m}(i) } d_{ij},   & \text{if } |N'_{d_m}(i)| \neq 0  \\
        -1,                                     & \text{otherwise}
    \end{cases},
    \quad
    \delta_i =
    \begin{cases}
        d_{i,nh_i}, & \text{if }  |N'_{d_m}(i)| \neq 0 \\
        +\infty,    & \text{otherwise}
    \end{cases},
\end{equation}
\noindent where $d_m= \max (\delta_o, \delta_c)$ and $N'_{d_m}(i) = \{ j : \rho_j > \rho_i, j \in N_{d_m}(i) \}$ is a subset of $N_{d_m}(i)$, where points have higher local densities than $\rho_i$.

% expand clusters
After $\rho$ and $\delta$ are calculated, points with density $\rho>\rho_c$ and large separation $\delta>\delta_c$ are promoted as cluster seeds, while points with density $\rho<\rho_c$ and large separation $\delta>\delta_o$ are demoted to outliers. For each point, there is a list of followers defined as:
\begin{equation} \label{eqn:algorithm:defineFollowers}
    F_i = \{j : nh_j=i \}.
\end{equation}
\noindent The lists of followers are built by registering the points which are neither seeds nor outliers to the follower lists of their nearest-highers. The cluster indices, associating a follower with a particular seed, are passed down from seeds through their chains of followers iteratively. Outliers and their descendant followers are guaranteed not to receive any cluster indices from seeds, which grants a noise rejection as shown in Fig.~\ref{fig:performance:outlierCuts}. The calculation of $\rho, \delta$ and the decision of seeds and outliers both support $n$-way parallelization, while the expansion of clusters can be done with $k$-way parallelization.
% theoretical complexity
Pseudocode of CLUE is included in Appendix~\ref{app:pseudocode}.
% For each of the $n$ points, CLUE computes $\rho$, $\delta$, list of followers and cluster index with a constant complexity granted by grid spatial index, resulting in $O(n)$ computational complexity. Besides, the space complexity is also $O(n)$ because CLUE only keeps a few algorithmic variables for each of $n$ points and does not rely on any $n\times n$ matrix.

\section{GPU Implementation}
\label{sec:implementation}

To parallelize CLUE on GPU, one GPU thread is assigned to each point, for a
total of $n$ threads, to construct spatial index, calculate $\rho$ and $\delta$,
promote (demote) seeds (outliers) and register points to the corresponding lists
of followers of their nearest-highers. Next, one thread is assigned to each
seed, for a total of $k$ threads, to expand clusters iteratively along chains of
followers. The block size of all kernels, which in practice does not have a
remarkable impact on the speed performance, is set to 1024. In the test in
Table~\ref{tbl:performance:breakdown}, changing the block size from 1024 to 256
on GPU leads to only about $0.14$~ms decrease in the sum of kernel execution
times. The details of parallelism for each kernel are listed in
Table~\ref{tbl:implementation:parallelism}. Since the results of a CLUE step are
required in the following steps, it is necessary to guarantee that all the
threads are synchronized before moving to the next stage. Therefore, each CLUE
step can be implemented as a separate kernel. To optimize the performance of
accessing the GPU global memory with coalescing, the points on all layers are
stored as a single structure-of-array (SoA), including information of their
layer numbers and 2D coordinates and weights. Thus points on all layers are
input into kernels in one shot.

\begin{table}[t]
    \renewcommand{\arraystretch}{1.25}
    % \small
    \centering
    \begin{tabular}{l|l|c|c}
        \hline
        Kernels                                  & parallelism    & total threads & block size \\
        \hline
        build fixed-grid spatial index           & 1 point/thread & n             & 1024 \\
        calculate local density                  & 1 point/thread & n             & 1024 \\
        calculate nearest-higher and separation  & 1 point/thread & n             & 1024 \\
        decide seeds/outliers, register followers& 1 point/thread & n             & 1024 \\
        expand clusters                          & 1 seed/thread  & k             & 1024 \\
        \hline
    \end{tabular}
    \caption{Kernels and Parallelism}
    \label{tbl:implementation:parallelism}
\end{table}

When parallelizing CLUE on GPU, thread conflicts to access and modify the same
memory address in global memory could happen in the following three cases:

\begin{enumerate}[i)]
    \item ~multiple points need to register to the same bin simultaneously;
    \item ~multiple points need to register to the list of seeds simultaneously;
    \item ~multiple points need to register as followers to the same point simultaneously.
\end{enumerate}

\noindent Therefore, atomic operations are necessary to avoid the race
conditions among threads in the global memory. During an atomic operation, a
thread is granted with an exclusive access to read from and write to a memory
location which is inaccessible to other concurrent threads until the atomic
operation finishes.

This inevitably leads to some microscopic serialization among threads in race.
The serialization in cases (i) and (iii) is negligible because bins are usually
small as well as the number of followers of a given point. In contrast,
serialization in case (ii) can be costly because the number of seeds $k$ is
large. This can cause delays in the execution of kernel responsible for seed
promotion. Since the atomic pushing back to the list of seeds is relatively fast
in GPU memory comparing to the data transportation between host and device, the
total execution time of CLUE still does not suffer significantly from the
serialization in case (ii). The speed performance is further discussed in
Section~\ref{sec:performance}.

\section{Performance Evaluation}
\label{sec:performance}

%---------------------------------------
% Clustering Results
%---------------------------------------
\subsection{Clustering results}
\label{sec:performance:clusteringResults}

% clustering on some non spherical topology
\begin{figure}[ht]
    \centering
    \includegraphics[clip, width=0.7\textwidth]{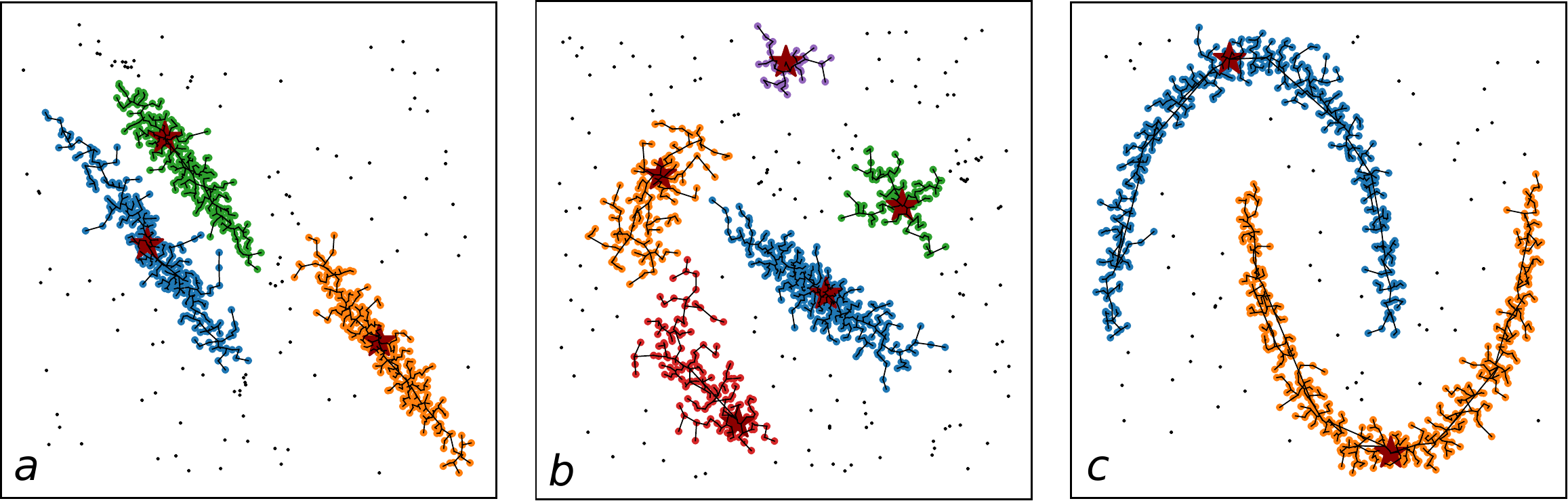}
    \caption{
    Examples of CLUE clustering on synthetic datasets. Each sample includes 1000
    2D points with the same weight generated from certain distributions,
    including uniform noise points. The color of points represent their cluster
    ids. Black points represent outliers detached from any clusters. The links
    between pairs of points illustrate the relationship between nearest-higher
    and follower. The red stars highlight the cluster seeds.  }
    \label{fig:performance:example}
\end{figure}

We demonstrate the clustering results of CLUE with a set of synthetic datasets,
shown in Fig.~\ref{fig:performance:example}. Each example has 1000 2D points and
includes spatially uniform noise points. The datasets in
Fig.~\ref{fig:performance:example} (a) and (c)  are from the scikit-learn
package~\cite{scikit-learn}. The dataset in Fig.~\ref{fig:performance:example}
(b) is taken from~\cite{rodriguez2014clustering}.
Fig~\ref{fig:performance:example} (a) and (b) include elliptical clusters and
Fig~\ref{fig:performance:example} (c) contains two parabolic arcs. CLUE
successfully detects density peaks in Figs.~\ref{fig:performance:example} (a),
(b), and (c).

\begin{figure}[ht]
    \centering
    \includegraphics[trim=3.5cm 0cm 3.5cm 0cm, clip,width=0.99\textwidth]{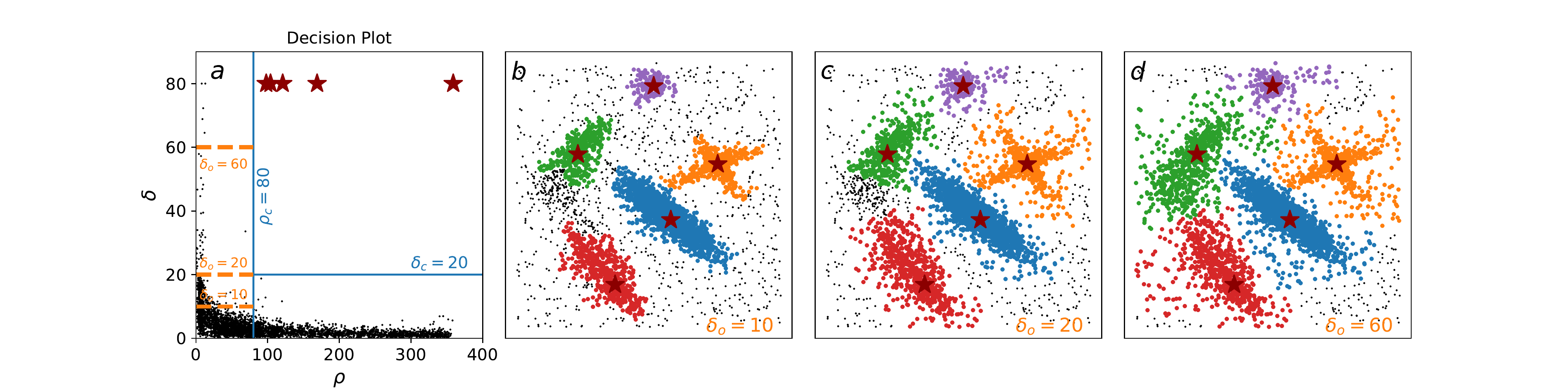}
    \caption{Noise rejection using different values of $\delta_o$. Noise is
    either an outlier or a descendant follower of an outlier. In this
    dataset~\cite{rodriguez2014clustering}, 4000 Points are distributed in
    $500\times500$ 2D square area. Figure~\ref{fig:performance:outlierCuts} (a)
    represents the decision plot on the $\rho-\delta$ plane, where fixed
    $\rho_c=80$ and $\delta_c=40$ values are shown as vertical and horizontal
    blue lines, respectively. Three different values of $\delta_o$ (10,20,60)
    are shown as orange dash lines. Figures~\ref{fig:performance:outlierCuts}
    (b), (c) and (d) show the results with $\delta_o=10,20,60$, respectively,
    illustrating how increasing $\delta_o$ loosens the continuity requirement
    and helps to demote outliers. The level of denoise should be chosen
    according to the user's needs.} \label{fig:performance:outlierCuts}
\end{figure}

In the induction principle of density-based clustering, the confidence of
assigning a low density point to a cluster is established by maintaining the
continuity of the cluster. Low density points with large separation should be
deprived of association to any clusters. CFSFDP uses a rather costly technique,
which calculates a boarder region of each cluster and defines core-halo points
in each cluster, to detach unreliable assignments from
clusters~\cite{rodriguez2014clustering}. In contrast, CLUE achieves this using
cuts on $\delta_o$ and $\rho_c$ while expanding a cluster, as described in
Section~\ref{sec:algorithm}. The example in
Fig.~\ref{fig:performance:outlierCuts} shows how cutting at different separation
values helps to demote outliers. Figure~\ref{fig:performance:outlierCuts} (a)
represents the decision plot on the $\rho-\delta$ plane. Points with density
below $\rho_c=80$, shown on the left side of the vertical blue line, could be
demoted as outliers if their $\delta$ are larger than a threshold. Once an
outlier is demoted, all its descendant followers are disallowed from attaching
to any clusters. While keeping $\rho_c=80$ fixed, the effect of using three
different values of $\delta_o$ (10, 20, 60), shown as orange dash lines in
Fig.~\ref{fig:performance:outlierCuts} (a), has been investigated. The
corresponding results are shown in Fig.~\ref{fig:performance:outlierCuts} (b),
(c) and (d), respectively.

%---------------------------------------
% Scaling
%---------------------------------------
\subsection{Execution time and scaling}
\label{sec:performance:executionTime}

\begin{figure}[ht!]
    \centering
    \includegraphics[width=0.9\textwidth]{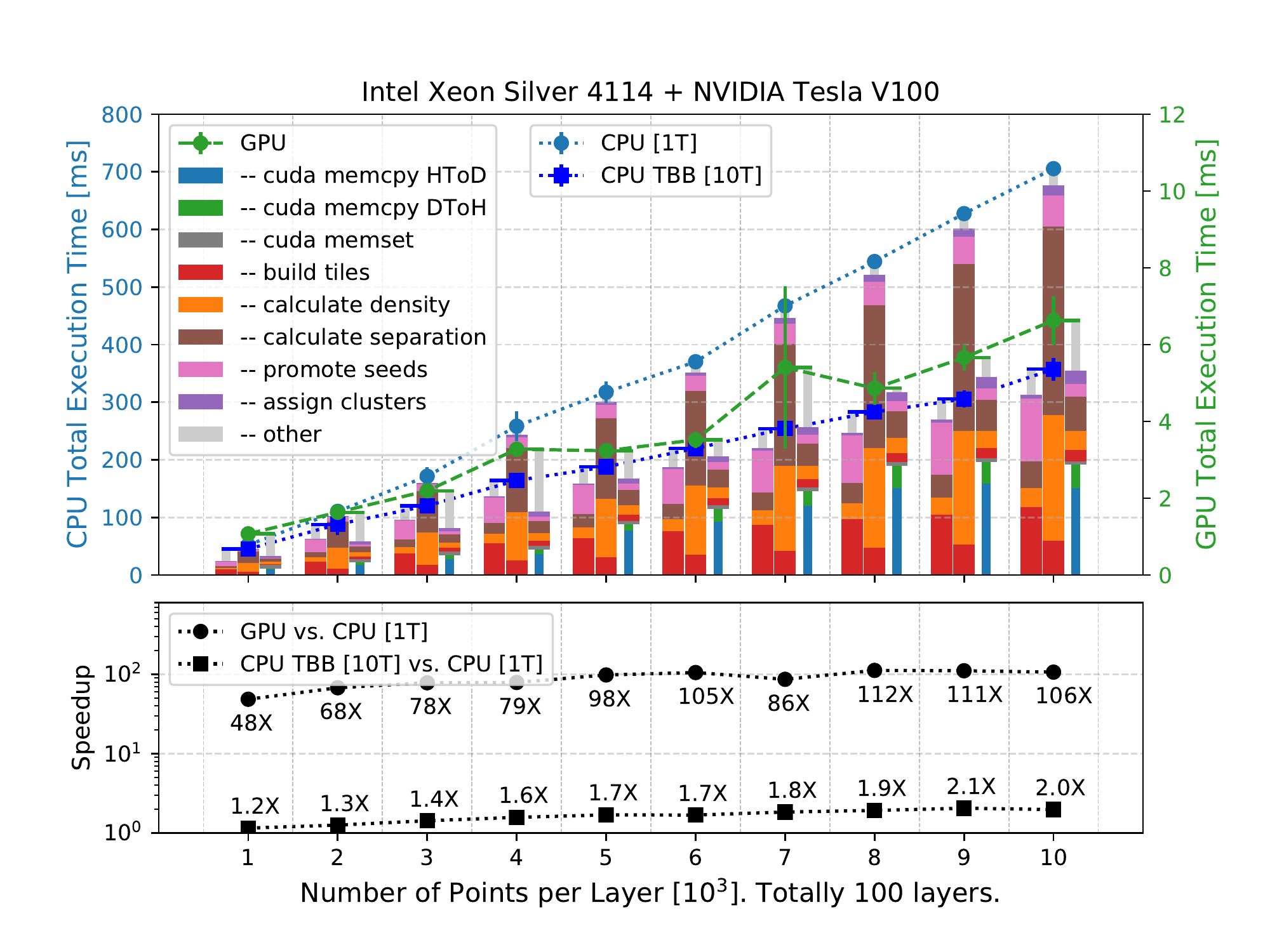}
    \caption{(\emph{Upper}) Execution time of CLUE on the single-threaded CPU,
    multi-threaded CPU with TBB and GPU scale linearly with number of input
    points, ranging from $10^5$ to $10^6$ in total. Execution time on
    single-threaded CPU is shown as blue circle dots and on 10 multi-threaded
    CPU with TBB is shown as blue square dots, while the time on GPU is shown as
    green circle dots. The stacked bars represent the decomposition of execution
    time. The green and blue narrower bars are latency for data traffic between
    host memory and device memory; wider bars represent time of essential CLUE
    steps; light grey narrower bars labelled as ``other'' are the difference
    between the total execution time and the sum of major CLUE steps (and major
    CUDA API calls if GPU). (\emph{Lower}) Comparing with the single-threaded
    CPU, the speed-up factors of the GPU range from 48 to 112,  while the
    speed-up factors of the multi-threaded CPU with TBB range from 1.2 to 2.0,
    which is less than the number of concurrent threads on CPU because of atomic
    pushing to the data containers discussed in
    Section~\ref{sec:implementation}. Table~\ref{tbl:performance:breakdown}
    shows the details of the decomposition of the execution time in the case of
    $10^4$ points per layer. }
    \label{fig:performance:executationTime}
\end{figure}

\begin{table}[ht!]
    \renewcommand{\arraystretch}{1.25}
    \tiny
    \centering
    % 10000
    \begin{tabular}{l|r@{}l|r@{}l|r@{}l}
    \hline
    CLUE Step                                 & \multicolumn{2}{c}{CPU [1T] (baseline)}         & \multicolumn{2}{c}{CPU TBB [10T]}                    & \multicolumn{2}{c}{GPU}  \\ \hline
    build fixed-grid spatial index            &  59.3 $\pm$&  ~1.6 ms       & 117.7 $\pm$&  ~6.4 ms ( 0.50x)        &   0.28 ms& ~(208.63x)       \\
    calculate local density                   & 218.4 $\pm$&  ~2.5 ms       &  33.7 $\pm$&  ~2.6 ms ( 6.48x)        &   0.51 ms& ~(430.57x)       \\
    calculate nearest-higher and separation   & 326.9 $\pm$&  ~2.9 ms       &  45.5 $\pm$&  ~2.5 ms ( 7.19x)        &   0.89 ms& ~(368.54x)       \\
    decide seeds/outliers, register followers &  54.4 $\pm$&  ~2.5 ms       & 109.4 $\pm$&  ~7.7 ms ( 0.50x)        &   0.34 ms& ~(162.38x)       \\
    expand clusters                           &  17.4 $\pm$&  ~1.5 ms       &   6.1 $\pm$&  ~1.3 ms ( 2.86x)        &   0.35 ms& ~( 49.74x)       \\ \hline
    cuda memcpy                               & --&                         & --&                                  &   2.87 ms&                   \\
    cuda memset                               & --&                         & --&                                  &   0.10 ms&                   \\
    other                                     &  29.1 $\pm$&  ~1.7 ms       &  44.9 $\pm$& ~15.7 ms                 &   1.30 ms&                  \\ \hline
    \textbf{TOTAL} (10000 points per layer)   & \textbf{705.49 $\pm$}&  ~\textbf{7.93 ms} & \textbf{357.24 $\pm$}& ~\textbf{19.68 ms ( 1.97x)} & \textbf{  6.63 $\pm$ 0.63 ms}& ~\textbf{(106.42x)}  \\
    \hline
    \end{tabular}

    \caption{Decomposition of CLUE execution time in the case of $10^4$ points
    per layer with 100 layers. The time of sub-processes on GPU is measured with
    NVIDIA profiler, while on CPU is measured with \texttt{std::chrono} timers
    in the C++ code. The uncertainties are the standard deviations of 200 trial
    runs of the same event (10000 trial runs if GPU). The uncertainties of
    sub-processes on GPU are negligible given that the maximum and minimum kernel execution time measured by NVIDIA Profiler are very close. With respect to the single-threaded CPU, the speed-up factors of the multi-threaded CPU with TBB and the GPU are given in the bracket. ``other" represents the difference between total execution time and the sum of the execution time of CLUE steps (and major CUDA API calls if GPU).}
    \label{tbl:performance:breakdown}
\end{table}

% testing dataset

We tested the computational performance of CLUE using a synthetic dataset that
resembles high occupancy events in high granularity calorimeters operated at
HL-LHC. The dataset represents a calorimeter with 100 sensor layers. A fixed
number of points on each layer are assigned a unit weight in such a way that the
density represents circular clusters of energy whose magnitude decreases
radially from the centre of the cluster according to a Gaussian distribution
with the standard deviation, $\sigma$, set to $3$~cm.  $5$\% of the points
represent noise distributed uniformly over the layers. When clustering with
CLUE, the bin size is set to $5$~cm comparable with the width of the clusters
and the algorithm parameters are set to $d_c=3 \text{ cm},\delta_o=\delta_c=5
\text{ cm},\rho_c=8$. To test CLUE's linear scaling, the number of points on
each layer is incremented from 1000 to 10000 in 10 equalling steps. A total of
100 layers are input to CLUE simultaneously which simulates the proposed CMS
HGCAL design~\cite{Collaboration:2293646}. Therefore the total number of points
in the test ranges from $10^5$ to $10^6$.
The linear scaling of execution time are validated in
Fig.~\ref{fig:performance:executationTime}.

The single-threaded version of the CLUE algorithm on CPU has been implemented in
C++, while the one on GPU has been implemented in C with
CUDA~\cite{nvidia2011nvidia}. The multi-threaded version of CLUE on CPU uses the
Thread Building Block (TBB) library~\cite{reinders2007intel} and has been
implemented using the Abstraction Library for Parallel Kernel Acceleration
(Alpaka)~\cite{zenker2016alpaka}. The test of the execution time is performed on
an Intel Xeon Silver 4114 CPU and NVIDIA Tesla V100 GPU connected by PCIe Gen-3
link. The time of each GPU kernel and CUDA API call is measured using the NVIDIA
profiler. The total execution time is averaged over 200 identical events (10000
identical events if GPU). Since CLUE is performed event by event and it is not
necessary to repeat memory allocation and release for each event when running on
GPU, we perform a one-time allocation of enough GPU memory before processing
events and a one-time GPU memory deallocation after finishing all events.
Therefore, the one-time \emph{cudaMalloc} and \emph{cudaFree} are not included
in the average execution time. Such exclusion is legit because the number of
events is extremely massive in high energy physics experiments and the execution
time of the one-time \emph{cudaMalloc} and \emph{cudaFree} reused by each
individual event is negligible.

In Fig.~\ref{fig:performance:executationTime} (\emph{upper}), the scaling of
CLUE is linear, consistent with the expectation. The execution time on the
single-threaded CPU, multi-threaded CPU with TBB and GPU increases linearly with
the total number of points. The stacked bars represent the decomposition of
execution time. In the decomposition, unique to the GPU implementation is the
latency of data transfer between host and device, which is shown as blue and
green narrower bars, while common to all the three implementations are the five
CLUE steps. Comparing with the single-threaded CPU, when building spatial index
and deciding seeds, shown as red and pink bars, the multi-threaded CPU using TBB
does not give a notable speed-up due to the implementation of atomic operations
in Alpaka~\cite{zenker2016alpaka} as discussed in
Section~\ref{sec:implementation}, while the GPU has a prominent outperformance
thanks to its larger parallelization scale. For the GPU case, the kernel of
seed-promotion in which serialization exists due to atomic appending of points
in the list of seeds, does not affect the total execution time significantly if
compared with other sub-processes. In the two most computing-intense steps,
calculating density and separation, there are no thread conflicts or inevitable
atomic operations. Therefore, both the multi-threaded CPU using TBB and the GPU
provide a significant speed-up. The details of the decomposition of execution
time in the case of $10^4$ points per layer are listed in
Table~\ref{tbl:performance:breakdown}.

Figure~\ref{fig:performance:executationTime} (\emph{lower}) shows the speed-up
factors. Compared to the single-threaded CPU, the CUDA implementation on GPU is
48-112 times faster, while the multi-threaded version using TBB via Alpaka with
10 threads on CPU is about 1.2-2.0 times faster. The speed-up factors are
constrained to be smaller than the number of concurrent threads because of the
atomic operations that introduce serialization. In
Table~\ref{tbl:performance:breakdown}, the speed-up factors of multi-threaded
CPU using TBB reduce to less than $1$ in the sub-process steps of building
spatial index and promoting seeds and registering followers, where atomic
operations happen and bottleneck the overall speed-up factor.

\section{Conclusion}
\label{sec:conclusion}

The clustering algorithm is an important part in the shower reconstruction of
high granularity calorimeters to identify hot regions of energy deposits. It is
required to be computationally linear with data scale $n$, independent from
prior knowledge of the number of clusters $k$ and conveniently parallelizable
when $n > k \gg m \equiv \frac{n}{k}$ in 2D. However, most of the well-known
algorithms do not simultaneously support linear scalability and easy
parallelization. CLUE is proposed to efficiently perform clustering tasks in
low-dimension space with $n > k \gg m$, including (and beyond) the applications
in high granularity calorimeters. The clustering time scales linearly with the
number of input hits in the range of multiplicity that is relevant for, e.g.,
the high granularity calorimeter of the CMS experiment at CERN. We evaluated the
performance of CLUE on synthetic data and demonstrated its capability of
non-spherical cluster shape with adjustable noise rejection. Furthermore, the
studies suggest that CLUE on GPU outperforms single thread CPU by more than an
order of magnitude within the data scale ranging from $n=10^5$ to $10^6$.

\bibliography{bibliography}

\section*{Acknowledgments}
The authors would like to thank the CMS and HGCAL colleagues for the many suggestions received in the development of this work. The authors would like to thank Vincenzo Innocente for the suggestions and guidance while developing the clustering algorithm. The authors would also like to thank Benjamin Kilian and George Adamov for their helpful discussion during the development of CLUE.
This material is based upon work supported by the U.S. Department of Energy, Office of Science, Office of Basic Energy Sciences Energy Frontier Research Centers program under Award Number DE-SP0035530, and the European project with CUP H92H18000110006, within the “Innovative research fellowships with industrial characterization” in the National Operational Program FSE-ERDF Research and Innovation 2014-2020, Axis I, Action I.1.

\section*{Code and Data Availability Statement}
The code and datasets for this study can be found at: \href{https://gitlab.cern.ch/kalos/clue/tree/V_01_20}{gitlab.cern.ch/kalos/clue}.

\newpage
\appendix

\section{Pseudocode}
\label{app:pseudocode}

Pseudocode of CLUE in serialized implementation. Parallelization is discussed in Section~\ref{sec:implementation}.

\begin{algorithm}[H]
% \SetAlgoLined
    \For{$i \in$ points}{
        $\rho_{[i]} = 0$ \\
        \For{$j \in \Omega_{d_c}(i)$ }{
            \If{$ dist(i,j) < d_c$}{
                $\rho_{[i]}$ += $w_{[j]}$
            }
        }
    }
\caption{calculate $\rho$}    
\end{algorithm}

\begin{algorithm}[H]
% \SetAlgoLined
    \For{$i \in$ points}{
        $\delta_{[i]} = +\infty$ \\
        $nh_{[i]} = -1$ \\
        \For{$j \in \Omega_{d_m}(i)$ }{
            \If{ $dist(i,j) < d_m$ \textbf{and} $\rho_{[j]} > \rho_{[i]}$}{
                \If {$dist(i,j) < \delta_{[i]} $}{
                    $nh_{[i]} = j$  \\
                    $\delta_{[i]} = d_{ij}$ \\
                }
            }
        }
    }
\caption{calculate $\delta$}    
\end{algorithm}

\begin{algorithm}[H]
% \SetAlgoLined
    k = 0\;
    stack = [] \;
    \For{i $\in$ points}{
        $isSeed = \rho_{[i]} > \rho_c \textbf{ and } \delta_{[i]} > \delta_c$ \\
        $isOutlier = \rho_{[i]} < \rho_c \textbf{ and } \delta_{[i]} > \delta_o$ \\
        \eIf {$isSeed$}{
            $clusterId_{[i]}$ = k \\
            k++ \\
            stack.pushback(i) \\
        } { \If{not $isOutlier$ }{
            $followers_{[nh_{[i]}]}.pushback(i)$
            }
        }
    }
    
    \While{stack.size $>$ 0}{
        i = stack.back \\
        stack.popback \\
        \For{$j \in followers_{[i]}$}{
            $clusterId_{[j]} = clusterId_{[i]}$ \\
            stack.pushback(j) \\
        }
    }
    
\caption{find seeds and outliers, assign clusters}
\label{algo:algorithm:assignClusters}
\end{algorithm}

% \newpage
% \section{Result on Toy Detector Dataset}
% \label{app:toyDetector}

% \begin{figure}[ht]
%     \centering
%     \includegraphics[width=0.49\textwidth]{figures/toyDetector/toyDetector_3000_5_10_5_8.png}
%     \includegraphics[width=0.49\textwidth]{figures/toyDetector/toyDetector_4000_5_10_5_8.png}
%     \includegraphics[width=0.49\textwidth]{figures/toyDetector/toyDetector_5000_5_10_5_8.png}
%     \includegraphics[width=0.49\textwidth]{figures/toyDetector/toyDetector_6000_5_10_5_8.png}
%     \caption{ Synthetic dataset used for the measurements of execution time in Section~\ref{sec:performance:executionTime}. Upper left, upper right, lower left, lower right are distribution of points on one layer with 3000, 4000, 5000, 6000 points respectively. }
%     \label{fig:app:toyDetector}
% \end{figure}

% \begin{figure}[ht]
%     \centering
%     \includegraphics[width=0.49\textwidth]{figures/toyDetector/toyDetector_7000_5_10_5_8.png}
%     \includegraphics[width=0.49\textwidth]{figures/toyDetector/toyDetector_8000_5_10_5_8.png}
%     \includegraphics[width=0.49\textwidth]{figures/toyDetector/toyDetector_9000_5_10_5_8.png}
%     \includegraphics[width=0.49\textwidth]{figures/toyDetector/toyDetector_10000_5_10_5_8.png}
%     \caption{ Result on Toy Detector Dataset. }
%     \label{fig:app:toyDetector}
% \end{figure}

\end{document}